\documentclass{PoS}

\title{Precision top pair production at hadron colliders}
\ShortTitle{Precision top production at hadron colliders}

\author{Micha\l{}  Czakon\\
        Institut f\"ur Theoretische Teilchenphysik und Kosmologie, RWTH Aachen University, D-52056 Aachen, Germany\\
        E-mail: \email{mczakon@physik.rwth-aachen.de}}
\author{\speaker{Alexander Mitov}\\
       Theory Division, CERN, CH-1211 Geneva 23, Switzerland\\
       E-mail: \email{alexander.mitov@cern.ch}}

\abstract{We demonstrate the impact of recent NNLO calculations on precision top quark phenomenology.}

\FullConference{{\rm Preprint numbers:} CERN-PH-TH/2012-358, TTK-13-03\\ \vskip 5mm
``Loops and Legs in Quantum Field Theory" 11th DESY Workshop on Elementary Particle Physics \\
		 April 15-20, 2012\\
		 Wernigerode, Germany
		 }

\begin{document}

\section{Introduction}

This writeup was inspired by three recent developments in collider physics. First, a Higgs-like boson was discovered at CERN \cite{:2012gk,:2012gu}, yet its precise identity remains unclear at present. Second, the otherwise widely appreciated connection between the top quark and Standard Model's Higgs boson was further strengthened in Refs.~\cite{Degrassi:2012ry,Mangano:2012mh}. Finally, new results in precision top pair production appeared \cite{Baernreuther:2012ws,Czakon:2012zr,Czakon:2012pz}, driving the precision in top physics to a quantitatively new level. In the following we give two examples that illustrate the power of such precision results.

\section{Tevatron: a comparison between alternative resummation approaches}\label{sec:tev}

Circa early 2012, the time this Workshop was held, the best theoretical precision in top pair production was achieved through either NNLL threshold resummation \cite{Beneke:2009rj,Czakon:2009zw,Ahrens:2010zv} on top of the NLO results or through approximate NNLO \cite{Beneke:2009ye} calculations based on truncation of the resummed NNLL results. A number of phenomenological studies appeared \cite{Ahrens:2010zv,Langenfeld:2009wd,Beneke:2010fm,Kidonakis:2010dk,Ahrens:2011mw,Ahrens:2011px,Kidonakis:2011ca,Beneke:2011mq,Cacciari:2011hy} with numerical predictions that showed a significant spread. 

The main reason behind such a spread was identified in Ref.~\cite{Cacciari:2011hy} as due to missing NNLO corrections that are not enhanced at threshold. Having the exact NNLO results for all partonic channels but the $gg$-initiated one, one can try to validate this assertion. As a test case we take the results of Refs.~\cite{Cacciari:2011hy} and \cite{Beneke:2011mq}, both including NNLL soft-gluon resummation matched through approximate NNLO. Both results resum directly the total inclusive cross-section. 
\footnote{It is possible \cite{Czakon:2009zw} to arrive at the resummed total inclusive cross-section by integration over the resummed differential distribution, as was done in  \cite{Ahrens:2010zv,Kidonakis:2010dk,Ahrens:2011mw,Ahrens:2011px} . Different power suppressed terms are generated in this case.}
There are two main differences between these two calculations: different resummation approaches (in $N$- and $x$-spaces, respectively) as well as different choices for the unknown two loop hard matching constants.

The completion of the exact NNLO calculation for the Tevatron \cite{Baernreuther:2012ws} allowed the extraction of the relevant two-loop hard matching constant and the fixing of all dominant two-loop power suppressed terms. Upon updating the results of Refs.~\cite{Cacciari:2011hy} and \cite{Beneke:2011mq} to promote them to full NNLO+NNLL accuracy (at the Tevatron) it was established that the spread of the predictions decreased dramatically, see  table~\ref{tab:table}. 
\begin{table}[ht]
\begin{center}
\begin{tabular}{| l | c | c |}
\hline
 & {\rm $N$-space}  & {\rm $x$-space} \\
\hline 
{\rm approximate NNLO + NNLL} & $6.72^{+0.24}_{-0.41}$ [{\rm pb}] & $7.22^{+0.29}_{-0.46}$ [{\rm pb}]\\ 
\hline 
{\rm exact~NNLO + NNLL} & $7.07^{+0.14}_{-0.23}$ [{\rm pb}]& $7.15^{+0.21}_{-0.20}$  [{\rm pb}]\\ 
\hline
\end{tabular}
\caption{\label{tab:table} Comparison between resummed predictions in $N$- and $x$-space matched to approximate or exact NNLO.}
\end{center}
\end{table}
Such a comparison can be easily performed with the help of the programs {\tt Top++} \cite{Czakon:2011xx} and {\tt TOPIXS} \cite{Beneke:2012wb}, respectively, and it demonstrates that indeed the spread in the numerical predictions is not (mainly) due to the different resummation formalisms. A similar comparison has been performed in Ref.~\cite{Beneke:2012wb}.
It will be very interesting to perform such a comparison also at the LHC, once the complete $gg$-initiated NNLO correction becomes available.

\section{Examining the quality of an approximation based on the high-energy limit of the cross-section}\label{high-energy}

The variety of NNLO approximations for the total inclusive top pair cross-section was extended in Ref.~\cite{Moch:2012mk} to include its high-energy limit, where the partonic energy $s$ is much larger than the mass of the top quark, i.e.  $\rho\equiv 4m^2/s\approx 0$. It is well understood \cite{Nason:1987xz,Catani:1990xk,Collins:1991ty,Catani:1990eg,Catani:1993ww,Catani:1994sq} that in this limit the cross-section exhibits logarithmic divergence: 
\begin{equation}
\sigma_{\rm tot} \approx c_1\ln(\rho) + c_0 + {\cal O}(\rho) \, .
\label{HElimit}
\end{equation}
The analytical result for the constant $c_1$ in all partonic channels can be found in Ref.~\cite{Ball:2001pq}.

In Ref.~\cite{Moch:2012mk} elegant arguments were given that allowed the authors to predict approximate numerical values for the constant $c_0$ in all partonic channels. It was noted there that the value of the constant is important for stabilizing the behavior of the high-energy approximation. Subsequently, the constants for the $q\bar q$ and $qg$ channels  were computed  in Refs.~\cite{Czakon:2012zr,Czakon:2012pz} and found to agree with the prediction of Ref.~\cite{Moch:2012mk}, within the numerical uncertainties. Two different applications of the high-energy approximation have emerged in the literature since then; we discuss them in the following.

\subsection{Application 1: Devising an approximation to the exact result}

The authors of Ref.~\cite{Moch:2012mk} argue that the high-energy limit of the cross-section, combined with its known threshold behavior, can be used to provide an approximation that is valid in the full kinematical region. In particular, they also argue that it can be used for improved phenomenological predictions for top pair production at hadron colliders. 

It was first pointed out in Ref.~\cite{CMS:alpha_s} that theoretical predictions including (or not) the approximation of Ref.~\cite{Moch:2012mk} differ about $7\%$ with respect to each other. The source of this difference was later investigated in Ref.~\cite{Czakon:2012pz} where the approximate result of Ref.~\cite{Moch:2012mk} was compared with the exact NNLO result in the $qg$ initiated reaction. The shift in the hadronic cross-section at the LHC at 7 and 8 TeV due to the $qg$ reaction alone was found to be about $5\%$. Such a difference is numerically very significant since the overall contribution of the $qg$ reaction at NNLO is only around $1\%$. 

The conclusion in Ref.~\cite{Czakon:2012pz} was that the shape of the approximation \cite{Moch:2012mk} in the mid-region is qualitatively different from the exact result. This region contributes significantly to the total cross-section for typical LHC c.m. energies due to the shape of the parton fluxes. Further arguments why such an approximation may not describe adequately the mid-range region can be deduced from the exactly known NLO result: it was shown in Ref.~\cite{Czakon:2008ii} that power expansions of the cross-section, around either the threshold or the high-energy limit are not well convergent and cannot describe the mid-region even after the inclusion of a number of powers. In the following we investigate the quality of the approximation of Ref.~\cite{Moch:2012mk} also in the $q\bar q$ reaction, which is now also known at NNLO \cite{Baernreuther:2012ws,Czakon:2012zr}. 
\begin{figure}[tbp]
\centering
\includegraphics[width=15cm]{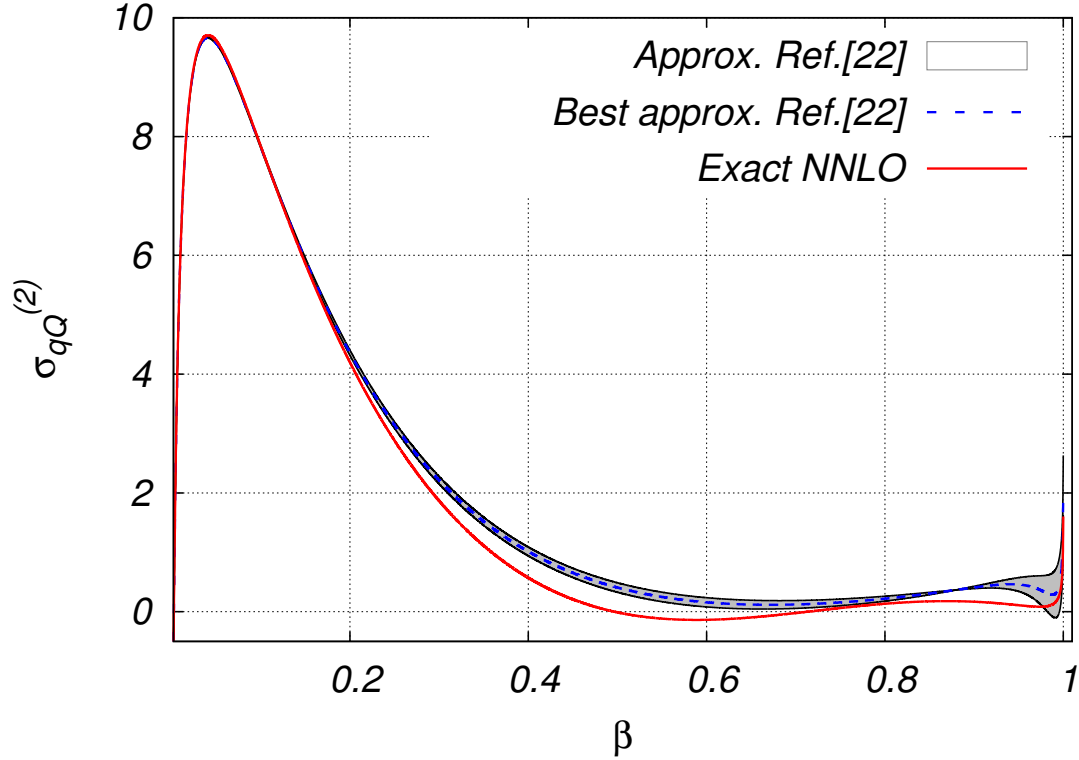}
\caption{\label{fig:partonic} Various NNLO corrections to the partonic cross-section: the exact NNLO result (red) and the approximation of Ref.~\cite{Moch:2012mk} (black band; best prediction is in blue). }
\end{figure}
\begin{figure}[tbp]
\centering
\hspace{-.5cm} 
\includegraphics[width=7.9cm]{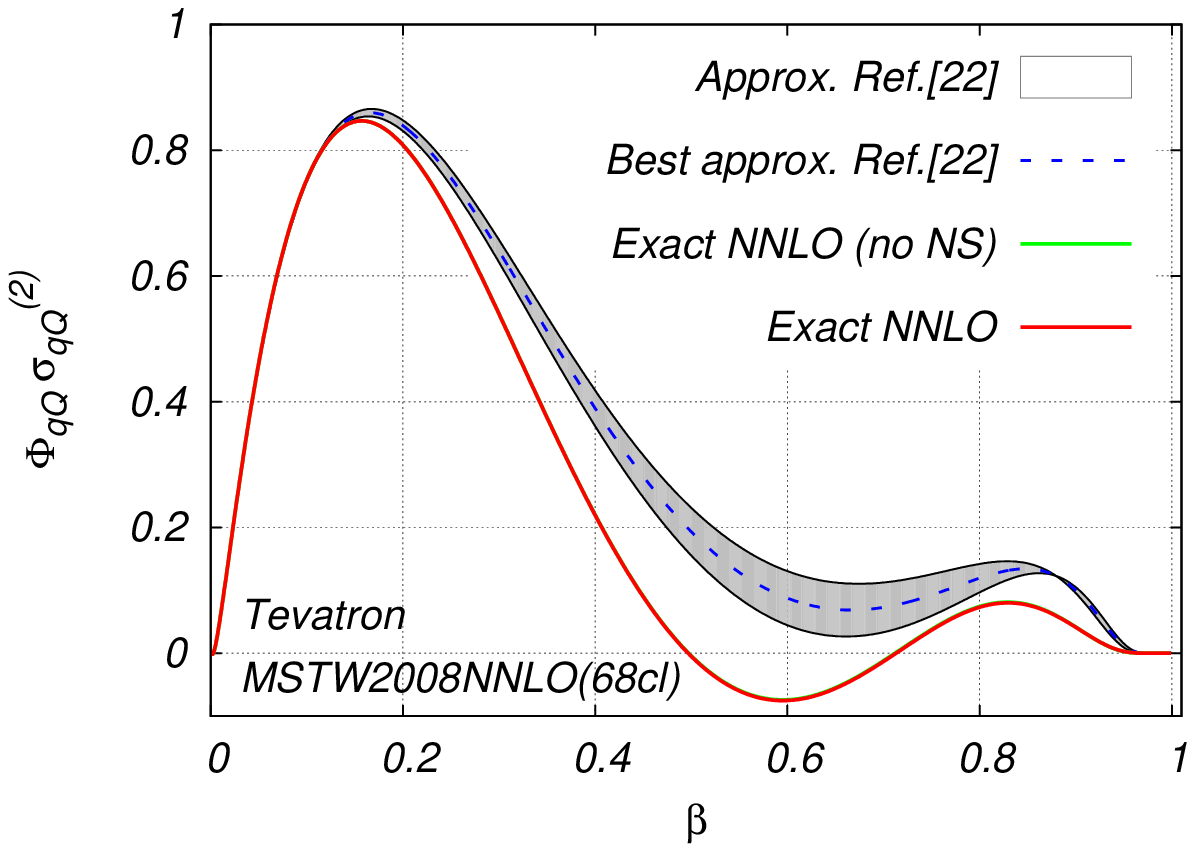}
\hspace{-0.5cm} 
\includegraphics[width=7.9cm]{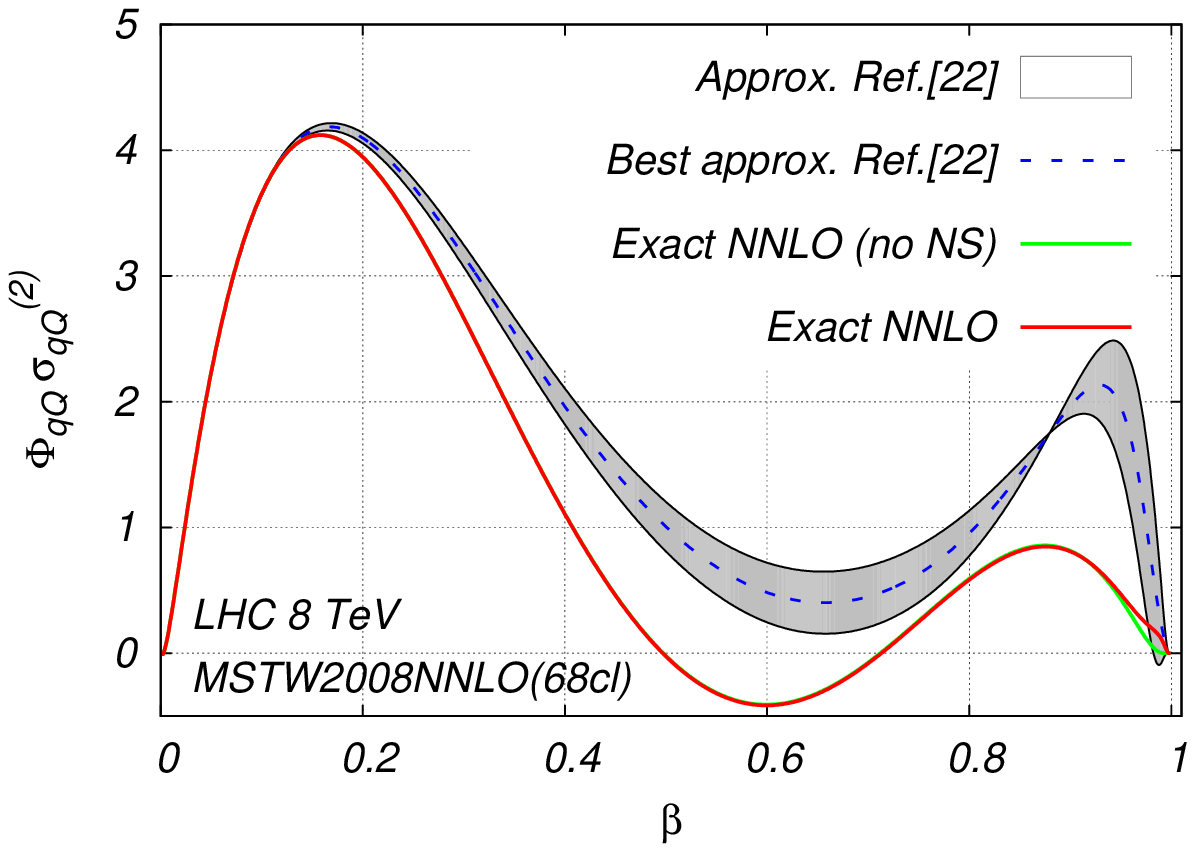}
\caption{\label{fig:fluxes} Various NNLO corrections to the partonic cross-section times $q\bar q$ partonic flux at Tevatron (left) and LHC 8 TeV (right): exact NNLO result (red), exact NNLO excluding the non-singlet correction (NS) \cite{Czakon:2012zr} that contributes all the high-energy rise (green), and the approximation of Ref.~\cite{Moch:2012mk} (black band; best prediction is in blue). }
\end{figure}

Similarly to the $qg$-initiated reaction \cite{Czakon:2012pz}, from fig.~\ref{fig:partonic} we conclude that in the mid-region the approximation of the partonic cross-section deviates noticeably with respect to the exact result. The significance of this difference becomes even more pronounced once the partonic cross-sections are multiplied by the respective partonic flux, see fig.~\ref{fig:fluxes}.  

It is also easy to check that the contribution to the total cross-section at the Tevatron due to the the difference between the exact NNLO and the best prediction of Ref.~\cite{Moch:2012mk} leads to a shift of around $0.14$ [pb] (a relative shift in the total cross-section of about 2\%), which is similar in size to the current scale variation at this collider.

\subsection{Application 2: Improving the limiting behavior of the full (numeric) result}

The leading high-energy behavior Eq.~(\ref{HElimit}) is utilized in Refs.~\cite{Czakon:2012pz,Czakon:2012zr} in the following way. The calculation of the NNLO partonic cross-sections \cite{Baernreuther:2012ws,Czakon:2012zr,Czakon:2012pz} is numeric, on a grid of finite number of points (typically around 80). This implies that one cannot calculate the partonic cross-section at, or close to, the two endpoints (threshold and high-energy limits) since both points are singular and therefore any numerical calculation will inevitably break there. It is in these regions, between the two kinematical endpoints and the lowest (highest) numerically computed point, where the knowledge of the asymptotic behavior proves to be very useful: one requires the fit of the numerical result to agree in both limits with the analytically known expansions, while the rest is covered by the numerical calculation which is reliable away from the two endpoints. 

As explained in Ref.~\cite{Czakon:2012pz} the error that one produces in this inherently numerical approach is very small and phenomenologically irrelevant for productions of top quarks at any near-future hadron collider. Applications at the LHC for very light quarks - like bottom - are likely also possible, but the errors have to be re-assessed based on the criterion given in Ref.~\cite{Czakon:2012pz}.

\section{Conclusions}

The level of precision in the description of top pair production at hadron colliders has increased dramatically in the last one year. The NNLO results for all partonic channels but the $gg$-initiated one have been calculated, and implemented in precision phenomenological studies. Comparing these exact result with various approximations we conclude that the mid-region behavior of the partonic cross-sections is very important and cannot be deduced from any available partial result or approximation.

\acknowledgments
We would like to thank the organizers for creating such a wonderful scientific atmosphere. We would also like to thank Sven Moch for very interesting discussion that triggered some of the studies presented in this proceedings. The work of M.C. was supported by the Heisenberg and by the Gottfried Wilhelm Leibniz programmes of the Deutsche Forschungsgemeinschaft, and by the DFG Sonderforschungsbereich/Transregio 9 "Computergest\"utzte Theoretische Teilchenphysik". The work of A.M. is supported by ERC grant 291377 "LHCtheory: Theoretical predictions and analyses of LHC physics: advancing the precision frontier".


\begin{thebibliography}{99}

%\cite{:2012gk}
\bibitem{:2012gk} 
  G.~Aad {\it et al.}  [ATLAS Collaboration],
  %``Observation of a new particle in the search for the Standard Model Higgs boson with the ATLAS detector at the LHC,''
  Phys.\ Lett.\ B
  [arXiv:1207.7214 [hep-ex]].
  %%CITATION = ARXIV:1207.7214;%%

%\cite{:2012gu}
\bibitem{:2012gu} 
  S.~Chatrchyan {\it et al.}  [CMS Collaboration],
  %``Observation of a new boson at a mass of 125 GeV with the CMS experiment at the LHC,''
  Phys.\ Lett.\ B
  [arXiv:1207.7235 [hep-ex]].
  %%CITATION = ARXIV:1207.7235;%%

%\cite{Degrassi:2012ry}
\bibitem{Degrassi:2012ry} 
  G.~Degrassi, S.~Di Vita, J.~Elias-Miro, J.~R.~Espinosa, G.~F.~Giudice, G.~Isidori and A.~Strumia,
  %``Higgs mass and vacuum stability in the Standard Model at NNLO,''
  JHEP {\bf 1208}, 098 (2012)
  [arXiv:1205.6497 [hep-ph]].
  %%CITATION = ARXIV:1205.6497;%%

%\cite{Mangano:2012mh}
\bibitem{Mangano:2012mh} 
  M.~L.~Mangano and J.~Rojo,
  %``Cross Section Ratios between different CM energies at the LHC: opportunities for precision measurements and BSM sensitivity,''
  JHEP {\bf 1208}, 010 (2012)
  [arXiv:1206.3557 [hep-ph]].
  %%CITATION = ARXIV:1206.3557;%%

%\cite{Baernreuther:2012ws}
\bibitem{Baernreuther:2012ws} 
  P.~Baernreuther, M.~Czakon and A.~Mitov,
  %``Percent Level Precision Physics at the Tevatron: First Genuine NNLO QCD Corrections to $q \bar{q} \to t \bar{t} + X$,''
  Phys.\ Rev.\ Lett.\  {\bf 109}, 132001 (2012)
  [arXiv:1204.5201 [hep-ph]].
  %%CITATION = ARXIV:1204.5201;%%

%\cite{Czakon:2012zr}
\bibitem{Czakon:2012zr} 
  M.~Czakon and A.~Mitov,
  %``NNLO corrections to top-pair production at hadron colliders: the all-fermionic scattering channels,''
  JHEP {\bf 1212}, 054 (2012)
  [arXiv:1207.0236 [hep-ph]].
  %%CITATION = ARXIV:1207.0236;%%  

%\cite{Czakon:2012pz}
\bibitem{Czakon:2012pz} 
  M.~Czakon and A.~Mitov,
  %``NNLO corrections to top pair production at hadron colliders: the quark-gluon reaction,''
  JHEP {\bf 1301}, 080 (2013)
  [arXiv:1210.6832 [hep-ph]].
  %%CITATION = ARXIV:1210.6832;%%

%\cite{Beneke:2009rj}
\bibitem{Beneke:2009rj} 
  M.~Beneke, P.~Falgari and C.~Schwinn,
  %``Soft radiation in heavy-particle pair production: All-order colour structure and two-loop anomalous dimension,''
  Nucl.\ Phys.\ B {\bf 828}, 69 (2010)
  [arXiv:0907.1443 [hep-ph]].
  %%CITATION = ARXIV:0907.1443;%%

%\cite{Czakon:2009zw}
\bibitem{Czakon:2009zw} 
  M.~Czakon, A.~Mitov and G.~F.~Sterman,
  %``Threshold Resummation for Top-Pair Hadroproduction to Next-to-Next-to-Leading Log,''
  Phys.\ Rev.\ D {\bf 80}, 074017 (2009)
  [arXiv:0907.1790 [hep-ph]].
  %%CITATION = ARXIV:0907.1790;%%

%\cite{Ahrens:2010zv}
\bibitem{Ahrens:2010zv}
  V.~Ahrens, A.~Ferroglia, M.~Neubert, B.~D.~Pecjak and L.~L.~Yang,
  %``Renormalization-Group Improved Predictions for Top-Quark Pair Production at
  %Hadron Colliders,''
  JHEP {\bf 1009}, 097 (2010)
  [arXiv:1003.5827 [hep-ph]].
  %%CITATION = JHEPA,1009,097;%%

%\cite{Beneke:2009ye}
\bibitem{Beneke:2009ye} 
%  M.~Beneke {\it et al.},
  M.~Beneke, M.~Czakon, P.~Falgari, A.~Mitov and C.~Schwinn,  
  %``Threshold expansion of the gg(qq-bar) ---> QQ-bar + X cross section at O(alpha(s)**4),''
  Phys.\ Lett.\ B {\bf 690}, 483 (2010)
  [arXiv:0911.5166 [hep-ph]].
  %%CITATION = ARXIV:0911.5166;%%

%\cite{Langenfeld:2009wd}
\bibitem{Langenfeld:2009wd} 
  U.~Langenfeld, S.~Moch and P.~Uwer,
  %``Measuring the running top-quark mass,''
  Phys.\ Rev.\ D {\bf 80}, 054009 (2009)
  [arXiv:0906.5273].% [hep-ph]].
  %%CITATION = ARXIV:0906.5273;%%

%\cite{Beneke:2010fm}
\bibitem{Beneke:2010fm}
  M.~Beneke, P.~Falgari, S.~Klein and C.~Schwinn,
  %``Threshold expansion of massive coloured particle cross sections,''
  Nucl.\ Phys.\ Proc.\ Suppl.\  {\bf 205-206} (2010) 20
  [arXiv:1009.4011 [hep-ph]].
  %%CITATION = NUPHZ,205-206,20;%%

%\cite{Kidonakis:2010dk}
\bibitem{Kidonakis:2010dk}
  N.~Kidonakis,
  %``Next-to-next-to-leading soft-gluon corrections for the top quark cross
  %section and transverse momentum distribution,'' 
  Phys.\ Rev.\  D {\bf 82} (2010) 114030
  [arXiv:1009.4935 [hep-ph]].
  %%CITATION = PHRVA,D82,114030;%%
    
%\cite{Ahrens:2011mw}
\bibitem{Ahrens:2011mw} 
  V.~Ahrens, A.~Ferroglia, M.~Neubert, B.~D.~Pecjak and L.~-L.~Yang,
  %``RG-improved single-particle inclusive cross sections and forward-backward asymmetry in $t\bar t$ production at hadron colliders,''
  JHEP {\bf 1109}, 070 (2011)
  [arXiv:1103.0550 [hep-ph]].
  %%CITATION = ARXIV:1103.0550;%%

%\cite{Ahrens:2011px}
\bibitem{Ahrens:2011px} 
  V.~Ahrens, A.~Ferroglia, M.~Neubert, B.~D.~Pecjak and L.~L.~Yang,
  %``Precision predictions for the t+t(bar) production cross section at hadron colliders,''
  Phys.\ Lett.\ B {\bf 703}, 135 (2011)
  [arXiv:1105.5824 [hep-ph]].
  %%CITATION = ARXIV:1105.5824;%%

%\cite{Kidonakis:2011ca}
\bibitem{Kidonakis:2011ca} 
  N.~Kidonakis and B.~D.~Pecjak,
  %``Top-quark production and QCD,''
  Eur.\ Phys.\ J.\ C {\bf 72}, 2084 (2012)
  [arXiv:1108.6063 [hep-ph]].
  %%CITATION = ARXIV:1108.6063;%%

%\cite{Beneke:2011mq}
\bibitem{Beneke:2011mq} 
  M.~Beneke, P.~Falgari, S.~Klein and C.~Schwinn,
  %``Hadronic top-quark pair production with NNLL threshold resummation,''
  Nucl.\ Phys.\ B {\bf 855}, 695 (2012)
  [arXiv:1109.1536 [hep-ph]].
  %%CITATION = ARXIV:1109.1536;%%

%\cite{Cacciari:2011hy}
\bibitem{Cacciari:2011hy} 
  M.~Cacciari, M.~Czakon, M.~Mangano, A.~Mitov and P.~Nason,
  %``Top-pair production at hadron colliders with next-to-next-to-leading logarithmic soft-gluon resummation,''
  Phys.\ Lett.\ B {\bf 710}, 612 (2012)
  [arXiv:1111.5869 [hep-ph]].
  %%CITATION = ARXIV:1111.5869;%%

%\cite{Czakon:2011xx}
\bibitem{Czakon:2011xx} 
  M.~Czakon and A.~Mitov,
  %``Top++: A Program for the Calculation of the Top-Pair Cross-Section at Hadron Colliders,''
  arXiv:1112.5675 [hep-ph].
  %%CITATION = ARXIV:1112.5675;%%
  
%\cite{Beneke:2012wb}
\bibitem{Beneke:2012wb} 
  M.~Beneke, P.~Falgari, S.~Klein, J.~Piclum, C.~Schwinn, M.~Ubiali and F.~Yan,
  %``Inclusive Top-Pair Production Phenomenology with TOPIXS,''
  JHEP {\bf 1207}, 194 (2012)
  [arXiv:1206.2454 [hep-ph]].
  %%CITATION = ARXIV:1206.2454;%%


%\cite{Moch:2012mk}
\bibitem{Moch:2012mk} 
  S.~Moch, P.~Uwer and A.~Vogt,
  %``On top-pair hadro-production at next-to-next-to-leading order,''
  Phys.\ Lett.\ B {\bf 714}, 48 (2012)
  [arXiv:1203.6282 [hep-ph]].
  %%CITATION = ARXIV:1203.6282;%%

%\cite{Nason:1987xz}
\bibitem{Nason:1987xz}
  P.~Nason, S.~Dawson and R.~K.~Ellis,
 % ``The Total Cross-Section for the Production of Heavy Quarks in Hadronic Collisions,''
  Nucl.\ Phys.\  B {\bf 303}, 607 (1988).
  %%CITATION = NUPHA,B303,607;%%
  
%\cite{Catani:1990xk}
\bibitem{Catani:1990xk} 
  S.~Catani, M.~Ciafaloni and F.~Hautmann,
%  ``Gluon contributions to small-x heavy flavor production,''
  Phys.\ Lett.\ B {\bf 242}, 97 (1990).
  %%CITATION = PHLTA,B242,97;%%

%\cite{Collins:1991ty}
\bibitem{Collins:1991ty} 
  J.~C.~Collins and R.~K.~Ellis,
  %``Heavy quark production in very high-energy hadron collisions,''
  Nucl.\ Phys.\ B {\bf 360}, 3 (1991).
  %%CITATION = NUPHA,B360,3;%%
  
%\cite{Catani:1990eg}
\bibitem{Catani:1990eg} 
  S.~Catani, M.~Ciafaloni and F.~Hautmann,
  %``High-energy factorization and small x heavy flavor production,''
  Nucl.\ Phys.\ B {\bf 366}, 135 (1991).
  %%CITATION = NUPHA,B366,135;%%

%\cite{Catani:1993ww}
\bibitem{Catani:1993ww}
 S.~Catani, M.~Ciafaloni and F.~Hautmann,
 %``High-energy factorization in QCD and minimal subtraction scheme,''
 Phys.\ Lett.\ B {\bf 307} (1993) 147.
 %%CITATION = PHLTA,B307,147;%%

%\cite{Catani:1994sq}
\bibitem{Catani:1994sq} 
  S.~Catani and F.~Hautmann,
  %``High-energy factorization and small x deep inelastic scattering beyond leading order,''
  Nucl.\ Phys.\ B {\bf 427}, 475 (1994)
  [hep-ph/9405388].
  %%CITATION = HEP-PH/9405388;%%

%\cite{Ball:2001pq}
\bibitem{Ball:2001pq} 
  R.~D.~Ball and R.~K.~Ellis,
  %``Heavy quark production at high-energy,''
  JHEP {\bf 0105}, 053 (2001)
  [hep-ph/0101199].
  %%CITATION = HEP-PH/0101199;%%

%\cite{CMS:alpha_s}
\bibitem{CMS:alpha_s} 
  [CMS collaboration] CMS note CMS-PAS-TOP-12-022.

%\cite{Czakon:2008ii}
\bibitem{Czakon:2008ii}
  M.~Czakon and A.~Mitov,
 % ``Inclusive Heavy Flavor Hadroproduction in NLO QCD: the Exact Analytic Result,''
  Nucl.\ Phys.\  B {\bf 824}, 111 (2010)
  [arXiv:0811.4119 [hep-ph]].
  %%CITATION = NUPHA,B824,111;%% 

\end{thebibliography}
\end{document}